\documentstyle[aps]{revtex}
\begin{document}
\draft
\twocolumn[\hsize\textwidth\columnwidth\hsize\csname@twocolumnfalse\endcsname

\title{A proposal on the possibility of detecting D-particles}
\author{Elias Gravanis and Nick E. Mavromatos } 
\address{Department of Physics, Theoretical Physics, King's College London,
Strand, London WC2R 2LS, U.K.}
\maketitle
\begin{abstract}
In a previous study we showed that D0-branes (particles) can 
operate as sources
of very-high-energy radiation, 
as a result of an unstable spacetime distortion that follows
a D0-particle/closed string state scattering. The 
effect can happen only if the energy of
the incident particle/closed string state exceeds a certain threshold, 
which is of order of 
the subsequently radiated energy. In this 
letter we speculate on the possibility of 
detection of the D-particles due to this phenomenon. The energies of the 
emitted
radiation range in a narrow window 
of size less than one order of magnitude. Observation-wise, 
this radiation 
will appear as an excess of photons in a narrow band in the spectrum
of high energy cosmic rays. 
From their energy we can then read off the value of the product 
of the string mass scale times 
the string coupling. 
We speculate on the possibility that 
high energy neutrinos from GRB's play the r\^ole of the necessary  high energy
flux that triggers the effect, by 
striking D-particles lying inside the mean free path of the photons
as measured from Earth. The possibilty of this effect operating as
a mechanism of GZK cutoff violation is also pointed out. 
\end{abstract} 

\vskip2pc]

\quad It has been argued in ref.\cite{recoil} that the recoil of 
a D-particle embedded in a four-dimensional spacetime, due to 
its scattering with a stringy
(high-energy) mode, results in the following metric: \begin{eqnarray} ds^2 =
\frac{{b'}^2 r^2}{t^2}dt^2 - \sum_{i=1}^{3} dx_i^2~, \qquad r^2 =
\sum_{i=1}^{3} x_i^2 \label{metricrecoil} \end{eqnarray}  where $b'(E)$ is the
momentum uncertainty of the recoiling $D$-particle, and $E$
denotes its kinetic energy. It is a property of $b'(E)$ to
decrease with increasing energy~\cite{szabo}. However, this energy
dependence is weak even for the high energies we consider here, and
thus $b'$ can be taken 
to be $b'\simeq 2g_s$, where $g_s$ is the string
coupling constant.

In a previous work by the authors~\cite{gravanis} it was shown that
this metric leads to the formation of a finite radius spherical
distortion of the initially flat spacetime  
surrounding the recoiling defect, that we call a ``bubble".
The metric exterior to the bubble, $r \ge 1/b'(E)$, is the flat Minkowski spacetime.
The interior of the bubble, on the other hand, contains
the excitations of an axion field (coming from the antisymmetric
tensor field of the string multiplet), and a tachyonic-like mode,
which expresses an instability of the bubble, of average 
lifetime~\footnote{In ref.~\cite{gravanis} we 
used the coordinate time $t'$ 
in the ``comoving frame'', which is related to the time $t$ in 
(\ref{metricrecoil}) 
by $dt'/l_s=dt/t$.  Note that, in the context of the recoil
formalism~\cite{szabo,kogan}, $t$ is the Minkowski time.
However, use of $t$ would require knowledge of the time moment at which 
the bubble was formed, and started decaying. Obviously, within 
a quantum string formalism such a process would occur within 
a (uncertainty) time  $\sim l_s$. Such issues  do not arise if one uses the 
``comoving'' time $t'$ instead, and this motivated its use in \cite{gravanis}.} 
\begin{eqnarray}\label{lifetime} 
   \tau \sim l_s~ e^{1/2g_s}
\end{eqnarray} 
where $l_s$ is the string length scale, and $g_s$ 
the string coupling. 

Bubble formation requires high-values of the incident energy
roughly of order
\begin{eqnarray}\label{threshold}
E^{\rm threshold} \sim \frac{g_s}{100}M_s
\end{eqnarray} 
where $M_s=1/l_s$ is the string mass scale.

The interior of the bubble has been found to have
a non-zero temperature~\cite{gravanis}
\begin{eqnarray}
     T_0 \sim g_sM_s
\end{eqnarray} 
due to the Rindler-accelerating nature of the distorted spacetime
around the recoiling D-particle. As a result, the bubble
radiates an energy of order
\begin{eqnarray}
 E_{\rm rad} \sim E_{\rm threshold}
\end{eqnarray} 
within its lifetime.
It should be remarked that this
amount of energy happens to be the same as the axion-field
thermal energy~\cite{gravanis}, thereby implying
that the D-partilce defect will not experience any mass loss.
Therefore our model involves D-particles that remain stable. 
This property is obviously necessary. In addition, assuming stability 
we restrict ourselves to a certain class of D-branes, whose 
other properties are also important for our purpose. We will return
to this later on.

From energy-momentum conservation, which
can be rigorously proven within our mathematical framework
of D-particle recoil~\cite{kogan,szabo},
we have:
\begin{eqnarray}\label{enermom}
&~&{\rm energy:}  \qquad  E_{\rm in} + M_D = M_D + \frac{1}{2}M_D u^2 + E_{\rm rad},
\nonumber \\
&~& {\rm momentum:} \qquad E_{\rm in} = M_D u + p
\end{eqnarray} 
where $M_D=M_s/g_s$ is the rest mass of the D-particle,
$u$ is its (non-relativistic) recoil velocity,
$E=\frac{1}{2}M_Du^2 $ is its kinetic energy,
$p$ denotes the total momentum of the emitted radiation from the bubble,
and we restricted ourselves for definiteness to the case
of a massless incident particle of energy $E_{\rm in}$.
From (\ref{enermom}) one can show that the total momentum $p$
satisfies the inequality
\begin{eqnarray}
    p \ge \frac{1}{2} E_{\rm rad}
\end{eqnarray} 
From this it is evident that the emitted number of photons,
during the bubble's life time,
is restricted to a few~\footnote{As a side remark
we observe that, if $E_{\rm in} $ is larger than
$(1+\sqrt{2})E_{\rm rad} \simeq 2.4E_{\rm rad}$, then
the total momentum of radiation is larger than its toral energy.
In such a case only a single photon is emitted
out of the bubble, and such an emission
is therefore
necessarily accompanied by another particle (which could be the incident
stringy mode itself).}.

We shall now make use of the above phenomenon to provide
a means of detection of the D0-particles. We will show that, under a few
assumptions, radiation from a physically acceptable population of D-particles
in the universe can be observed. Then, an excess of photons somewhere
in the observed spectrum of high energy cosmic rays, can be interpreted as
coming from this effect and, through that, from the branes. 
Since high energy photons
have a mean free path smaller than 100 Mpc, 
we shall consider here D-particles
located inside a volume of radius of order of 10 Mpc, around Earth.

Consider a D-particle striken by a current of
highly-energetic weakly-interacting particles,
which can travel undisturbed at great distances.
We assume now that the energy of the particle
is higher than the threshold energy (\ref{threshold})
for the formation of the bubble. The highest 
energy neutrinos from gamma-ray bursters 
(GRB's)~\cite{neutrino} can provide 
the high energy flux we need, 
as in the bursters there is a relatively significant production of 
particles whose energies 
can possibly be of 
order $10^{-2}g_sM_s$, where 
$g_s$ is assumed smaller than unity, for the validity of the 
string perturbation expansion. 
Our interest in neutrinos arises from the fact that these particles 
can travel
undisturbed over very large distances, creating significant 
uniform and isotropic flux. According to contemporary models~\cite{neutrino}, 
these neutrinos 
come with appreciable flux only up to energies
$10^{18}$eV, so our first assumption will be that the threshold
energy must be bounded from above by such a value. This, in turn, implies 
that 
the string length scale falls within the experimental sensitivity, 
if it is at most (c.f. (\ref{threshold}))
\begin{eqnarray}\label{stringscale}
  M_s \sim \frac{10^{11}~GeV}{g_s}
\end{eqnarray} 
Then, a few photons will be emitted, with an energy spectrum
in a narrow window below $E_{rad}$, whose highest value has been assumed to be
about $10^{18}$eV~\footnote{Note that, if one uses 
$g_s\sim 10^{-2}$, which is 
small, as needed for
the validity of our string perturbation expansion, but not unjustifiably 
small,
we observe   
that the smallest detectable string length, in this approach,
is $l_s \sim 10^{-27} {\rm cm}$, in which case 
the bubble's lifetime 
(\ref{lifetime}) is $\sim 10^{-16}$ sec.}. 

We now turn to an estimation of the distribution 
of D-particles in the universe which can
produce sufficient rates of photons on Earth.
We assume a {\it uniform} and {\it isotrpopic}
flux $J_{\rm he}$ for the highly-energetic particles,
and a concentration of $D$-particles $n_D$ in a region of
linear dimension $\ell$. Then, it is evident that the
flux of the incident particles, which will strike the
$D$-particles, is $J_{\rm he}(1-e^{-\ell \sigma n_D})$
where $\sigma$ is the effective cross section of the
particle/D-brane scattering :
\begin{eqnarray}\label{cross}
  \sigma \sim l_s^{2} = (\frac{10^{11}}{g_s}GeV)^{-2} = g_s^{2}10^{-50} {\rm cm}^2.
\end{eqnarray} 
Note that the string length scale $l_s$ determines the 
natural size of the D-brane as seen by an incoming string.
For each scattering event there will be a few
highly-energetic photons emitted due to the mechanism
described in ref. \cite{gravanis} and reviewed above.
This implies that, in order of magnitude,
the total number of emitted photons is about the same
as that of incident particles. Assuming that $\ell \sigma n_D $ is small,
it follows that the flux of the emitted photons will be roughly of order
$J_{he}\ell\sigma n_D$. Restricting ourselves from now on to the value
$g_s=10^{-2}$, we obtain  $\sigma \sim 10^{-54}
{\rm cm^{2}}$.

The emerging photons of the first particle/D-brane scattering
carry energies less than the threshold energy,
hence, when such photons strike another D-particle will be captured.
If $\ell \sigma n_D $ is small,
the final photon flux will still be given by
\begin{eqnarray}\label{fluxgamma}
   J_\gamma = J_{\rm he}\ell \sigma n_D
\end{eqnarray} 
since in this case 
$e^{-\ell \sigma n_D} \simeq 1 $.

As already mentioned, a good candidate for the 
highly energetic particles with energies above 
the threshold (\ref{threshold}), 
are neutrinos emitted from the 
GRB's. 
According to studies of the last few years~\cite{neutrino}, neutrinos of such 
energies can be
emitted from the fireball due to the muon decay that follows
the pion production, due to the interaction of fireball protons with
afterglow photons. The energy of neutrinos of appreciable
flux is about two orders of magnitude less than that of 
protons, which is supposed to be the highest energy of the  
ultra-high-energy cosmic rays (UHECR). Then,  
we expect to have netrinos of flux of order, 
roughly, $10^{-9} {\rm GeV}/{\rm cm^{2}sec}$ at energies $10^{18}$ 
to $10^{19}$eV, 
depending on the energy and neutrino species~\cite{neutrino}.

If one assumes a uniform distribution of the D-particle defects
inside large regions in space, then a simple calculation shows that,
in order to explain the observed events, one needs a
phenomenologically unacceptable total mass of D-particles.
It is, therefore, necessary to 
assume that the latter form collections of very high 
density, whose concentration in space, however, is low.
We next assume that such collections can produce radiation flux of order
of that of the high energy cosmic rays,
$J_{\rm CR} \sim 10^{-20}/{\rm cm^{2}~sec~sr}$, and
that the observed UHECR flux is of the same order
as the photon flux (\ref{fluxgamma}). If there are $N$ such ensembles 
of $D$-particles
in a volume around 
Earth, of average radius in the order of 10 Mpc, one has
\begin{eqnarray} \label{relformula}
&~& J_{\rm CR} \sim  NJ_\gamma \sim NJ_{\rm he} \ell \sigma n_D =
Nc n_{\rm he} \ell \sigma n_D\sim \nonumber \\ 
&~& \sim N c n_{he}\sigma N_D/\ell^{2}
\end{eqnarray} 
where $N_D$ denotes the average number of branes in one of 
these D-brane-collections,
$c$ is the speed of light in vacuo,
and $n_{\rm he}$ is the concentration
of the high-energy particles, assumed to be neutrinos from
GRB's . Their flux, given above, implies a number density of high energy
particles $n_{he} \sim 10^{-28}/{\rm cm}^{3}$.
 Then,
\begin{eqnarray} \label{nepin}N N_D \sim \ell^{2}10^{48}/{\rm cm}^{2}. \end{eqnarray}  
Using the fact that $M_D=10^{-9}{\rm gr}$ for the given string scale and 
coupling, one obtains 
the total D-particle mass contained in the 10Mpc-volume as:
\begin{eqnarray} \label{totalmass}M_{total} \sim 10^{6} \ell^{2}/{\rm cm}^{2}~M_\odot \end{eqnarray} 
These
formulae show that in the limit of very small $\ell$ the total
number and mass of branes needed can be very small. We have then 
to determine the lower limit of the linear dimension of the brane
collections $\ell$. First we must take into account 
a few other requirements.
As we have stressed earlier, it is important that the scattering rate of
high energy particles passing through the brane ensembles
is small, so that 
\begin{equation} 
\ell \sigma n_D\sim \sigma N_D/ \ell^{2} \ll 1.
\label{ineq}
\end{equation} 
Also, in order to be able to treat the D-particle collections as distinct
spots of scattering, whose internal density is unrelated to their
concentration in space, we must have $n_D \gg N/(10 ~{\rm Mpc-volume})$. 
Finally ,
the Schwarzschild radius has to be much smaller than $\ell$, so
that one is far away from the black hole limit. Then, it turns out that 
the strongest condition 
is (\ref{ineq}), and the last two are 
then trivially satisfied.
Since there is no independent lower limit for
$N_D$, we may take for concreteness $N_D \sim 10$ assuming
that such a configuration forms a stable bound state of D-particles, 
an issue that we will discuss later on.
Setting $\ell$ to be ten orders of magnitude larger than the saturated 
limit, we then get
$\ell\sim 10^{-16}$ cm, which gives a value for the total number of
brane ensembles in the 10 Mpc-volume $N\sim 10^{16}$. These imply
a total D-brane mass in that volume of order 
$10^{-26} M_\odot \sim 10^{4} ~{\rm kgr}$. These numbers are quite sensitive in
changes of the parameters, but there are large regions of the
parameter space where the results are reasonable . The total mass
can be very small and hence it does not contribute
to the dark matter, although it would be a very good candidate for it, 
given that branes are ``bright'' only when very high energetic particles
are passing nearby.

The conclusion of the above, rather indicative, results is 
that a relatively
small number of D-particles is needed to produce an observable effect, if
our assumptions are valid. Then, an observed excess of photons in a certain 
energy channel of the high energy cosmic rays 
would imply the existence of D-particles, thereby 
fixing the value of the product $g_sM_s$ of the fundamental
string parameters. The above estimates provide
information only about a portion of the total population of D-particles
in the universe, assuming they exist, which contributes 
to an observable radiation.

It should be remarked that the above considerations 
were based on the assumption that the maximum energy 
of the incident particles (neutrinos or other species) 
is of order $10^{19}$ eV. Relaxing this, by allowing 
the maximum energies to be higher, e.g. of order 
$10^{21}$ eV~\cite{neutrino},
would imply that the above scenario could also provide~\cite{gravanis} 
a mechanism for production of UHECR beyond the Greisen-Zatsepin-Kuzmin
(GZK) cut-off~\cite{gzkcutoff}. For this to happen, of course, 
the location of the brane defects must lie within the respective
mean free paths. However, we must note that 
these considerations 
cannot offer an explanation of the complete set of the observed UHECR 
events~\cite{gzkobs}. The so-produced UHECR events, if any,  
constitute only a contribution to this set, 
given that our mechanism produces only 
photons, with energies in a rather narrow interval. 

For the validity of the above scenario it is crucial 
that the populations of the D-particle defects 
are {\it stable}. Naively, since the D-particles 
are very massive,
one expects that at large concentrations, such as the ones above,
they will collapse to form
black holes or at least to produce very strong gravitational fields,
which would jeopardize our scenario. 
The problem of preventing such a collapse
is equivalent to that of stabilizing D-branes, which
at present is an important open issue.
One expects that, in general, 
such stability mechanisms would 
imply a certain distance scale among 
the constituent D-particles in a collection, 
as well as restrictions
on the number of branes in the ensemble and probably 
on the total number of branes in the universe, something 
which most likely cannot be explained by the dynamics of their
interactions alone.

However, there are recent attempts in the string literature
in which -under certain circumstances- one has a {\it no force
condition} among the D-particles. Usually this is the
case of sufficient spacetime supersymmetries, where the
D-particle states are viewed as specific states, saturating the
so-called
Bogomoln\'yi-Prasad-Sommerfield (BPS) bound~\cite{dual}.
However, it has been pointed out~\cite{dual2}
that one may obtain {\it stable} non BPS D-brane states
by considering appropriate combinations of BPS branes,
in such a way that spacetime supersymmetry is not preserved.
From our point of view we are interested in such stable non-BPS
non-supersymmetric D-particles.

Such states
can be viewed as solutions of certain string theories, which are
connected with the phenomenologically relevant string theories,
like Heterotic String, assumed to be living on
the four-dimensional spacetime
(after appropriate compactification),
by virtue of certain {\it
duality symmetries}~\cite{dual,dual2}.
One such theory is type IIB string theory.
In general, D-particle/D-particle scattering in type IIB string theory
has been studied in the literature~\cite{typeiib}. Although the
construction of \cite{typeiib}
refers to a specific string model involving a particular orbifold
compactification~\cite{dual3}, which may not be necessarily
realized in our situation, however we find it generic enough
so as to consider it as a prototype for the D-particle
stability mechanism
we need here.

The main idea behind the stability mechanism via the no force condition
in this model for D-particles
lies on the fact that there are additional vector interactions,
$U_D(1)$,
associated with the specific way of
constructing the non-BPS state.
We shall not explain the details here, but we refer the interested
reader to the relevant literature~\cite{dual2,typeiib,dual3}.
Such $U_D(1)$ interactions should not be confused with observable interactions
of ordinary matter in our construction. The latter may be assumed {\it neutral}
under the $U_D(1)$, which thus characterizes only the D-particles
in the ensemble, which are then 
assumed `electrically' charged under this $U_D(1)$, and in principle
may have both positive and negative charges. The positive
charges characterize, say, the D-particles, whilst the negative charges
the anti-D particles,
denoted by ${\overline D}$.
The $U_D(1)$  interaction is therefore
{\it repulsive} among D-particles (or ${\overline D}$-particle), 
and {\it attractive} amond D-particle-${\overline D}$-particle.

Once we admit both kinds of charges
there  arises the issue as
to how ``polarized'' D-particle collections, characterized by
a significant 
excess of (positive or negative)  $U_D(1)$ charge, have been formed
in the Galaxies today. This
has not been resolved as yet, neither
will be
the topic of the present work. For our purposes here
we merely conjecture that, somewhow, after the
Big Bang, such polarized regions emerged inside ordinary matter,
in such a way that the overall net $U_D(1)$ charge of the
Universe
is zero. An alternative scenario would be
that ${\overline D}$-particles behaved in similar
way as antiparticles in ordinary matter, and hence the Universe
today consists in its overwhelming majority of D-particles
only. This, would then trivially solve
the above-mentioned problem of ``polarization''.

In general, the no force condition may be a property valid at
all distances. This is the case of supersymmetric D-particles that saturate
the BPS bound. However,
in the orbifold construction of \cite{typeiib}, the no force condition
among the non BPS D-particles
is found to occur at large scales $r$, as
compared with the string length $l_s$, where notably (target space) effective
low-energy string action methods are 
applicable. This feature 
is exclusive of a critical-radius orbifold compactification.

An important point to remark is that the construction of
\cite{typeiib} ignored higher-string loop effects.
The incorporation of such effects results in general
in the destruction of the no force condition~\cite{loops},
although under some circumstances it may be made valid up
to one loop. The effects
of string-loop resummation cannot be answered at present, and
hence the issue of the no force condition at a {\it non-perturnbative}
string level is still open.

\section*{Acknowledgements}

The work of E.G. is supported
by a King's College London Research Studentship (KRS).
N.E.M. wishes to thank H. Hofer (ETH, Zurich and CERN) 
for his interest and support.

\end{document}